\documentstyle[preprint,prl,aps]{revtex}
\begin{document}
\draft

\title{Reentrant Melting of Soliton Lattice Phase in Bilayer Quantum Hall System}

\author{S.~Park$^1$, K.~Moon$^{1,}$\cite{byline}, C.~Ahn$^2$, J.~Yeo$^3$,
C. Rim$^4$, and B.H.~Lee$^5$}
\address{$^1$Department of Physics and IPAP, Yonsei University, 
Seoul 120-749, Korea\\ 
$^2$Department of Physics, Ewha Womans University, Seoul 120-750, Korea\\
$^3$Department of Physics, Konkuk University, Seoul 143-701, Korea\\
$^4$Department of
Physics, Chonbuk National University, Chonju 561-756, Korea\\
$^5$Department of Physics, Sogang University, C.P.O. Box 1142,
Seoul 100-611, Korea}

\date{\today}

\maketitle
\begin{abstract}
At large parallel magnetic field $B_\parallel$, the ground state of bilayer quantum 
Hall system forms uniform soliton lattice phase. 
The soliton lattice will melt due to the proliferation  
of unbound dislocations at certain finite temperature 
leading to the Kosterlitz-Thouless (KT) melting.
We calculate the KT phase boundary by numerically solving the newly developed 
set of Bethe ansatz equations, which fully
take into account the thermal fluctuations of soliton walls.
We predict that within certain ranges of $B_\parallel$,   
the soliton lattice will melt at $T_{\rm KT}$. Interestingly enough, 
as temperature decreases, it melts at certain temperature lower than $T_{\rm KT}$
exhibiting the reentrant behaviour of 
the soliton liquid phase. 
\end{abstract}


\narrowtext
When the interlayer spacing $d$ is comparable to the mean particle distance, 
the bilayer quantum Hall system at total filling factor $\nu_{\rm tot}=1$
can exhibit the quantum Hall effect due to the strong interlayer 
correlations\cite{murphy,exp,moon}.
In the presence of large parallel magnetic field $B_\parallel$, 
the ground state 
of the system is known to form a uniform soliton lattice (SL) phase 
made of the  
periodic array of the phase solitons of the field variable $\theta ({\bf r})$, 
which represents the relative phase difference
between electrons in two layers\cite{moon}.
Using the isospin language \cite{moon}, the effective 
energy functional for the bilayer system can be written by 
\begin{equation}
E (\theta)=\int d^2r~\Biggl\{ \frac{1}{2}~\rho_s\vert {\bf\nabla}\theta\vert^2 -
\frac{t}{2\pi\ell^2}~\cos{[\theta ({\bf r}) + Qx]}\Biggr\},
\label{PTmodel}
\end{equation}
where $\rho_s$ is the isospin stiffness, $t=t_0 e^{-Q^2\ell^2/4}$ 
the interlayer tunneling amplitude,
$\ell = (\hbar c / |e| B_{\perp})^{1/2}$, and $Q=2\pi /L_\parallel$ with 
$L_\parallel = \Phi_0/B_\parallel d$, which defines the length associated 
with one flux quantum $\Phi_0$ enclosed between two layers.
We set the magnetic length $\ell=1$. 
This continuum model is valid, when the field $\theta ({\bf r})$ varies
smoothly over the lengths $\ell, L_\parallel$\cite{comment1}.
With the increase of $B_\parallel$, the system exhibits a quantum phase 
transition from commensurate (C) to incommensurate (I) phase at 
$Q_c = (2/\pi)(2t/\pi \rho_s)^{1/2}$ at $T=0$\cite{moon,pokrovsky}. 
The incommensurate phase at $Q > Q_c$ describes the uniform SL phase 
at zero temperature. 

In the paper, we study the thermodynamics of the SL 
phase through a mapping of the 2D statistical mechanical model  
in Eq. (\ref{PTmodel}) to 
the 1D quantum sine-Gordon (QSG) Hamiltonian. When one neglects the 
compactness of the field variable $\theta$, which will be 
reasonably valid for $T < (\pi/2)\rho_s$,  
the 1D QSG model is exactly soluble by the Bethe ansatz method.
The Bethe ansatz solution provides the thermodynamic CI phase
boundary $Q_c (T)$.
Depending on the values of 
$C=t/(32\pi \rho_s)$, the critical value $Q_c (T)$ will increase (decrease)
with temperature for $C>1\,\, (C<1)$.
Within the incommensurate phase, the ground state of the system is the 
uniform SL phase at $T=0$.
As $T$ increases to about $(\pi/2)\rho_s$, the compactness of 
the angle variable $\theta$ will 
become important, which introduces dislocations to the system leading to the 
KT melting of the SL phase. 
Recently Hanna {\em et al.} have studied the melting of the SL  
phase by analyzing the elastic moduli based on the rigidity of the zero 
temperature ground state\cite{hanna,coppersmith}. 
However, the renormalizaion of the elastic moduli due to thermal 
fluctuations of soliton walls has not been taken into account. 
Furthermore, as one approaches to the CI phase boundary, the soliton 
density will vanish.  
In order to understand the melting of the SL phase
near the CI phase boundary, it is crucial to include the  
thermal fluctuations and the soliton density variations as well. 

In the paper, we develop a new set of Bethe ansatz equations,
which fully take into account both the thermal fluctuations of 
the soliton walls and the density variations.  
By numerically solving the Bethe ansatz equations, we have calculated both 
the compression modulus $\kappa_{xx}$ and the shear modulus $\kappa_{yy}$  
of the SL phase. By making an asymptotic expansion near the CI phase boundary, 
we have shown that $\kappa_{xx}$ goes like $(2\tau/M)^{1/2}
\left( Q-Q_c (T)\right)^{1/2}\rho_s$ and  
$\kappa_{yy} \cong (8 M \tau^3)^{1/2}\left( Q-Q_c (T)\right)^{-1/2}
\rho_s$, where $M$ is the soliton mass which varies with $\tau$ and 
$\tau=T/(8\pi\rho_s)$. This asymptotic behaviour explicitly confirms the 
predictions made by Coppersmith {\em et al.}\cite{coppersmith}. 
Based on the elastic moduli thus obtained, $T_{\rm KT}$ is calculated 
and compared with   
the zero temperature estimates\cite{hanna,read}.  
We predict the following reentrant behaviour of the soliton liquid phase.
At certain values of $Q$ slightly below $Q_c (0)$,   
the system initially stays at the C phase. As $T$ increases, it makes  
a transition to the soliton liquid phase. With the further increase of
$T$, the soliton liquid solidifies to the SL phase due to the rapid increase 
of the soliton density, which subsequently melts reentering 
the soliton liquid phase.  

Following the effective energy functional of Eq. (\ref{PTmodel}),  
the low temperature thermodynamics of the system  
can be defined by the statistical partition function 
${\cal Z}=\int[{\cal D}\theta]e^{-E(\theta)/T}$.
One can map the statistical ensemble summation into quantum transfer
matrix by identifying ${\cal Z}=\lim_{R\rightarrow \infty} 
{\rm Tr}\left[e^{-R{\hat H}}\right]$, 
\begin{equation}
{\hat H}=\int dx\left[
{T\over{2\rho_s}}\Pi^2+{\rho_s\over{2T}}(\partial_x\theta)^2
-{t \over {2\pi T}} \cos\left(\theta + Qx 
\right)\right].
\end{equation}
If one neglects the compactness of the angle variable $\theta ({\bf r})$, 
the canonical conjugate momentum $\Pi ({\bf r})$ can be easily integrated 
out leading to the corresponding classical partition function.  
We make the following change of variables $\phi ({\bf r})=\sqrt{\rho_s/T} 
\left[\theta ({\bf r}) + Qx\right]$, which leads to 
the quantum Hamiltonian of the 1D sine-Gordon model with external 
$U(1)$ coupling, 
\begin{equation}
{\hat H}=\int dx\left[   
{1\over{2}}\Pi^2+{1\over{2}}(\partial_x\phi)^2
-A{\beta\over{2\pi}}\partial_x\phi+2\mu\cos\left(\beta
\phi\right) 
+{1\over{2T}}\rho_s Q^2\right].
\end{equation}
Here $\mu={t/(4\pi T)},\qquad \beta=\sqrt{{T/{\rho_s}}}, 
{\rm and} \qquad   
A={2\pi Q\rho_s/T}$. The compactness of $\theta ({\bf r})$ 
would restrict $\Pi ({\bf r})$ to the integer values.    
Hence the ground state energy ${\cal E}$ of the 1D  
QSG model corresponds to ${\cal F}/T$, where ${\cal F}$ is the 
free energy of the 2D statistical mechanical system.

The ground state of the 1D QSG model is exactly 
soluble by the Bethe ansatz method. 
There is a competition between the finite soliton mass $M$  
which prefers the commensurate phase and the external field $A$ coupled to
the topological charge which favors the incommensurate phase.
This is a `quantum' version of the CI transition\cite{perbak}, where 
the soliton mass includes full `quantum' fluctuation effects.
With external $U(1)$ field, a soliton can be created with energy cost
$\Delta {\cal E}= M\cosh\Theta-A$.
The soliton mass $M$ is given by
\begin{equation}
M={2\Gamma\left({p\over{2}}\right)\over{\sqrt{\pi}
\Gamma\left({p+1\over{2}}\right)}}
\left[\pi\mu{\Gamma\left({1\over{p+1}}\right)\over{
\Gamma\left({p\over{p+1}}\right)}}\right]^{{1\over{2}}(p+1)},
\end{equation}
where $p=T/(8\pi\rho_s-T)$ and $\Theta$ represents  
the rapidity of the soliton\cite{zamolochikov}.

For $A<M$, the ground state will be the vacuum: the
commensurate phase. However for $A>M$, a nontrivial vacuum will
arise, since the energy cost to create a soliton can be negative. 
In this case, the ground state will be described by the soliton 
condensations: The SL phase. 
The CI transition will occur at $A=M$.
By equating the two quantities, 
one can obtain the exact CI phase boundary $Q_c (T)$  
\begin{equation}
Q_c (T)={8\over{\sqrt{\pi}}}{\tau\Gamma\left({\tau\over{2(1-\tau)}}\right)
\over{\Gamma\left({1\over{2(1-\tau)}}\right)}}
\left[C{\Gamma(1-\tau)\over{\Gamma(1+\tau)}}\right]^{{1\over{2(1-\tau)}}},
\end{equation}
where we have neglected the $Q$-dependence of $t$, since the continuum model 
is only valid for $Q\le 1$. The parameter $C$ is given by    
${t/(4 T_0)}$.
The reduced temperature variable $\tau$ is between $0$ and $1$. 
As $\tau\to 0$, one can reproduce the classical limit: 
$Q_c (0)={16\sqrt{C}/{\pi}}$.


Now we want to study the melting of the SL phase by numerically solving the
Bethe ansatz equations.  
Using the fact that the exact scattering matrix is known  
for the 1D QSG model, one can calculate the ground state energy and 
the soliton density ${\bar n}_s (Q,T)$ as a function of $Q$ and 
$T$\cite{zamolochikov}.
The Bethe ansatz equation is given by 
\begin{equation}
2\pi\rho(\Theta)=M\cosh\Theta+\int_{-B}^{B}d\Theta'\varphi(\Theta-\Theta')
\rho(\Theta'),
\label{Bethe1}
\end{equation}
where $\rho(\Theta)$ is the density of solitons between $\Theta$ and
$\Theta+d\Theta$  per unit length and the 
integral kernel $\varphi (\Theta)$ is given by  
\begin{equation}
{\varphi}(\Theta)=\int_{-\infty}^{\infty} d\omega \, e^{i\omega\Theta} 
{\sinh\left({\pi(p-1)\omega\over{2}}\right)\over{
2\cosh\left({\pi\omega\over{2}}\right)\sinh\left({\pi p\omega\over{2}}\right)
}}.
\end{equation}
In terms of the density function, one can calculate the ground state
energy as follows 
\begin{equation}
\Delta {\cal E}=\int_{-B}^{B}d\Theta (M\cosh\Theta-A)\rho(\Theta), 
\label{Bethe2} 
\end{equation}
where the ground state is determined by imposing the condition: 
$\partial \Delta{\cal E}/\partial B =0 $.  
By making an asymptotic expansion of Eq. (\ref{Bethe1}) near the CI phase
boundary, one can confirm that the soliton density exhibits a power-law
behaviour: ${\bar n}_s \cong (2M/\tau)^{1/2} 
\left( Q-Q_c (T)\right)^{1/2}/(2\pi)$\cite{pokrovsky}.
%

As $T$ increases to about $(\pi/2)\rho_s$, 
the compactness of the variable 
$\theta ({\bf r})$ becomes important, which introduces topological defects 
(dislocations) into the system leading to the KT melting of the SL phase
at temperatures much below the CI transition temperature.
Although the 1D QSG Hamiltonian does not involve dislocation excitations, 
one can get a reasonable estimate for $T_{\rm KT}$ by analyzing the elastic 
moduli of the SL at finite temperature.   
The elastic moduli $\kappa_{ij}$ are defined
as follows\cite{hanna}
\begin{equation}
\kappa_{ij}={\partial^2 
\over {\partial Q_s^i \partial Q_s^j}}{\cal F},
\end{equation}
where $Q_s^x$ is given by $2\pi/L_s$, where $L_s$ is the lattice constant
of the SL phase along the ${\hat x}$ direction, $Q_s^y = \tan\phi\, Q_s^x$ with
$\phi$ the tilt angle of the soliton walls, and $i,j=x,y$.
In order to obtain the elastic moduli of the system, one needs to 
calculate the energy of the system by varying the soliton density 
from the ground state value {\em and/or} tilting the solitons 
from the vertical orientations. 
We first calculate the compression modulus $\kappa_{xx}$. 
By shifting the fermi momentum from $B$ to $B + \epsilon$,
one can increase (decrease) the soliton density for positive (negative) 
values of $\epsilon$. The soliton density $\rho (\Theta)$ is a function 
of both $\epsilon$ and $\Theta$.  
For small $\epsilon$, one can expand the soliton density  
up to quadratic order in $\epsilon$
\begin{equation}
\rho (\Theta) \cong {\bar \rho} (\Theta) + \epsilon \rho_b^\prime 
(\Theta)
+{\epsilon^2 \over 2} \rho_b'' (\Theta),
\end{equation}
where ${\bar \rho} (\Theta)$ is the soliton density profile of 
the ground state, 
$\rho_b^\prime = \partial \rho/\partial 
\epsilon |_{\epsilon=0}$, and $\rho_b'' = \partial^2 \rho /\partial \epsilon^2
|_{\epsilon=0}$. 
At the lowest order of $\epsilon$, we obtain 
the usual Bethe 
ansatz equation for the ground state as shown in Eq.(\ref{Bethe1}) with the
condition $\partial \Delta{\cal E}/\partial \epsilon = 0$.
Up to the order of $\epsilon^2$, we obtain the following two equations
\begin{equation}
2\pi\rho_b^\prime (\Theta) = [\varphi(\Theta-B) +  
\varphi(\Theta+B)] {\bar \rho} (B)
+\int_{-B}^{B} d\Theta^\prime \varphi (\Theta-\Theta^\prime) 
\rho_b^\prime (\Theta^\prime),   
\end{equation}
\begin{eqnarray}
2\pi\rho_b'' (\Theta) &=& {\partial \over {\partial \Theta}}
\left[ \varphi(\Theta+B) -  \varphi(\Theta-B)\right] 
{\bar \rho} (B) 
+ [\varphi(\Theta-B) +  
\varphi(\Theta+B) ] \left[ {\partial {\bar \rho} (B)
\over {\partial B}} +  2 \rho_b^\prime (B) \right] \nonumber\\ 
&+&\int_{-B}^{B} d\Theta^\prime \varphi (\Theta-\Theta^\prime) 
\rho_b'' (\Theta^\prime).
\end{eqnarray}
The energy of the system is given as follows:  
$\Delta{\cal E} = \Delta {\bar {\cal E}} + {\epsilon^2 \over 2} 
\chi_b$, where $\Delta {\bar {\cal E}}$ is the ground state energy and 
$\chi_b$ is given by   
\begin{equation}
\chi_b = 2 M  \sinh B {\bar \rho} (B) + 2 (M \cosh B - A)
\left[ {\partial {\bar \rho} (B)
\over {\partial B}} +  2 \rho_b^\prime (B)\right]
+\int_{-B}^{B} d\Theta (M\cosh\Theta - A)  
\rho_b'' (\Theta). 
\end{equation}
The $Q_s^x$ can be calculated as follows:     
$Q_s^x = 2\pi\int_{-B-\epsilon}^{B+\epsilon} d\Theta 
\rho (\Theta) \cong {\bar Q}_s^x  
 + \alpha_b \epsilon + {\cal O}(\epsilon^2)$,
where ${\bar Q}_s^x = 2\pi\int_{-B}^{B} d\Theta 
{\bar \rho} (\Theta)$ is the ground state value and
$\alpha_b=4\pi{\bar\rho} (B) + 2\pi \int_{-B}^{B} d\Theta 
\rho_b^\prime (\Theta)$.
The compression modulus $\kappa_{xx}$ is given as follows:  
$\partial^2 {\cal F}/{\partial Q_s^x}^2 = T\chi_b / \alpha_b^2$.

Now we calculate the shear modulus $\kappa_{yy}$. To obtain the shear modulus,
one needs to fix $Q_s^x$ and vary $Q_s^y$ alone.
Global shift of the rapidity will not only tilt the soliton walls but also
change the soliton density along the ${\hat x}$-direction.
In order to vary $Q_s^y$ alone, one needs to change the fermi momentum
$B$ to $B + \epsilon$ and rapidity by $\eta$ simultaneously along  
the following trajectory: 
$\epsilon = - {\alpha}_{\rm T}\eta^2$ for some constant ${\alpha}_{\rm T}$
\cite{KOSEF_long}. 
By making a perturbative expansion in $\eta$, we obtain the following two   
Bethe ansatz equations up to $\eta^2$,  
\begin{equation}
2\pi\rho_s^\prime (\Theta) = [\varphi(\Theta-B) -  
\varphi(\Theta+B)] {\bar \rho} (B)
+\int_{-B}^{B} d\Theta^\prime \varphi (\Theta-\Theta^\prime) 
\rho_s^\prime (\Theta^\prime), 
\end{equation}
\begin{eqnarray}
2\pi\rho_s'' (\Theta) &=& {\partial \over {\partial \Theta}}
\left[ \varphi(\Theta+B) -  \varphi (\Theta-B) \right] 
{\bar \rho} (B) 
+ \left( \varphi(\Theta-B) +  
\varphi(\Theta+B)\right) \left[ {\partial {\bar \rho} (B)
\over {\partial B}} +  2 \rho_s^\prime (B)\right] \nonumber\\
&+&\int_{-B}^{B} d\Theta^\prime \varphi (\Theta-\Theta^\prime) 
\rho_s'' (\Theta^\prime), 
\end{eqnarray}
where one can easily notice that $\rho_s^\prime (\Theta)$ is an 
odd function of $\Theta$ and $\rho_s'' (\Theta)$ an even function. 
The energy of the system along the trajectory is given by 
\begin{equation}
\Delta{\cal E} = \Delta {\bar{\cal E}} 
+ {\eta^2 \over 2} \chi_s,
\end{equation}
where $\chi_s$ is given by 
\begin{equation}
\chi_s = 2M\sinh B {\bar \rho} (B) + 2 (M\cosh B - A)
\left[ {\partial {\bar \rho} (B)
\over {\partial B}} +  2 \rho_s^\prime (B)\right] 
+ \int_{-B}^{B} d\Theta (M\cosh\Theta - A) \rho_s'' (\Theta).  
\end{equation}
Since  
$Q_s^y = \tan \phi \,{\bar Q}_s^x  
=\eta\,{\bar Q}_s^x$, the shear modulus $\kappa_{yy}$ is given by
the following relation:  
$\partial^2 {\cal F}/{\partial Q_s^y}^2 = T\chi_s / ({\bar Q}_s^x)^2$.
The off-diagonal elements of $\kappa_{ij}$ are zero. 
Using the asymptotic expansion, one can show that near the CI phase boundary,   
$\kappa_{xx}$ goes like $(2\tau/M)^{1/2} 
\left( Q-Q_c (T) \right)^{1/2}\rho_s$, and 
$\kappa_{yy}\cong (8M\tau^3)^{1/2} 
\left( Q-Q_c (T) \right)^{-1/2}\rho_s$\cite{KOSEF_long}.  

In Murphy {\em et al.}'s experiment, for equally populated layers,  
the experimental values of the various parameters are given as follows:  
$\rho_s \cong 0.35 K, t_0 \cong 1.2 K, \ell \cong 126 \AA$, 
and $d\cong 200 \AA$\cite{murphy}.     
For the above sample, the parameter $C$ can be estimated to be about $0.033$.   
In the inset of Fig. (1), the open circles represent $\kappa_{xx}$ and the open 
squares $\kappa_{yy}$ at fixed value of $Q\cong 0.919$, which obey the correct
asymptotic behaviour near the CI phase boundary. 
We have confirmed that both $\kappa_{xx}$ and $\kappa_{yy}$ approach to $\rho_s$ 
at large $Q$\cite{KOSEF_long}.
The KT transition temperature $T_{\rm KT}$ can be estimated by solving 
the following equation: $k_B T/(\pi/2) \rho_s =  
[\kappa_{xx}(T)\kappa_{yy}(T)]^{1/2}/\rho_s$\cite{read}.  
In Fig. (1), the closed circles represent   
$(\kappa_{xx}(T)\kappa_{yy}(T))^{1/2}/(16\rho_s)$ as a function of  
$\tau$. The intersections with the solid line with slope $1$ locate the positions  
of $T_{\rm KT}$. 
It can be shown that for $Q_{\rm min} < Q < Q_c (0)$,  
the equations have two solutions, where $Q_{\rm min} \cong 0.9, Q_c (0) \cong 0.927$. 
In Fig. (2), we plot the KT melting temperature as a function of $Q$. 
The closed circles  
represent the KT phase boundary from our Bethe ansatz calculation, 
the open squares from Hanna {\em et al.} based on the zero temperature 
value of the elastic moduli, and the solid curve the CI phase boundary.
For $Q_{\rm min} < Q < Q_c (0)$, the system initially stays at the C phase.
As $T$ increases, it makes a transition to the soliton liquid phase. 
With the further increase of $T$, the soliton liquid becomes the SL phase due to the
rapid increase of the soliton density upon entering 
the CI phase boundary.  
Subsequently it melts exhibiting the reentrant behaviour of the soliton liquid phase.  
We have confirmed that at large $Q$, $T_{\rm KT}$ approaches to $(\pi/2) \rho_s$
\cite{KOSEF_long}.  
Read argues that at $T=0$, quantum fluctuations are not important, 
since the domain `sheets' are marginally rough\cite{read,fisher}. 
However at finite $T$, thermal length scale $\xi_\beta = (\hbar v / k_B T)$
becomes finite. Hence near the CI phase boundary {\em and/or} KT phase  
boundary, the large distance thermal fluctuations will be still important. 

To summarize, we have studied the melting of the SL phase by
numerically solving the newly developed set of Bethe ansatz equations,
which fully take into account both the thermal fluctuations of 
the soliton walls and the density variations. 
Based on the elastic moduli thus obtained, $T_{\rm KT}$ is calculated.
We predict that the system will exhibit the reentrant behaviour of the 
soliton liquid phase within certain ranges of $Q$. 
Our proposed KT phase boundary can be experimentally observed by 
measuring the longitudinal resistance of the high purity bilayer sample.
Towards the completion of the paper, we became aware that the preprint 
{\em cond-mat}/0201343 also studied the bilayer quantum Hall system based on 
the exact solution of the 1D sine-Gordon model. In our paper, we mainly 
focus on the role of the compactness of the angle variables 
and explicitly calculate the KT phase boundary.  

K.M. acknowledges C. Hanna for useful discussions and allowing 
his data for the publication in our paper.  
This work was supported by grant No. R01-1999-00018 from the
interdisciplinary Research program of the KOSEF. 
K.M. acknowledges the financial support by Yonsei University 
Research Fund of 2000.

\begin{figure}
\caption{Determination of the KT melting temperature.  
The closed circles represent  
$(\kappa_{xx}(T)\kappa_{yy}(T))^{1/2}/(16\rho_s)$ as a function of
$\tau$. The intersections with the solid line with slope $1$ locate the positions
of $T_{\rm KT}$.
In the inset, the open circles represent the compression modulus $\kappa_{xx}$
as a function of $\tau$ at fixed $Q=0.919$ and the open squares
the shear modulus $\kappa_{yy}$.}
\label{fig1}
\end{figure}

\begin{figure}
\caption{The KT melting temperature as a function of $Q$.  
The KT phase boundary is shown   
and compared with the result of Hanna {\em et al.}[6],  
where $\tau_{\rm KT}=16\tau$.
The closed circles are from our Bethe ansatz result and the open squares
from Hanna {\em et al.}. 
The solid line represents the CI phase boundary.} 
\label{fig2}
\end{figure}


\begin{references}
\bibitem{comment1} At large parallel magnetic field such that $L_\parallel 
\cong \ell$, one may as well use the Hartree-Fock approximation, which
extends the validity regime to much shorter length scale.
\end{references}
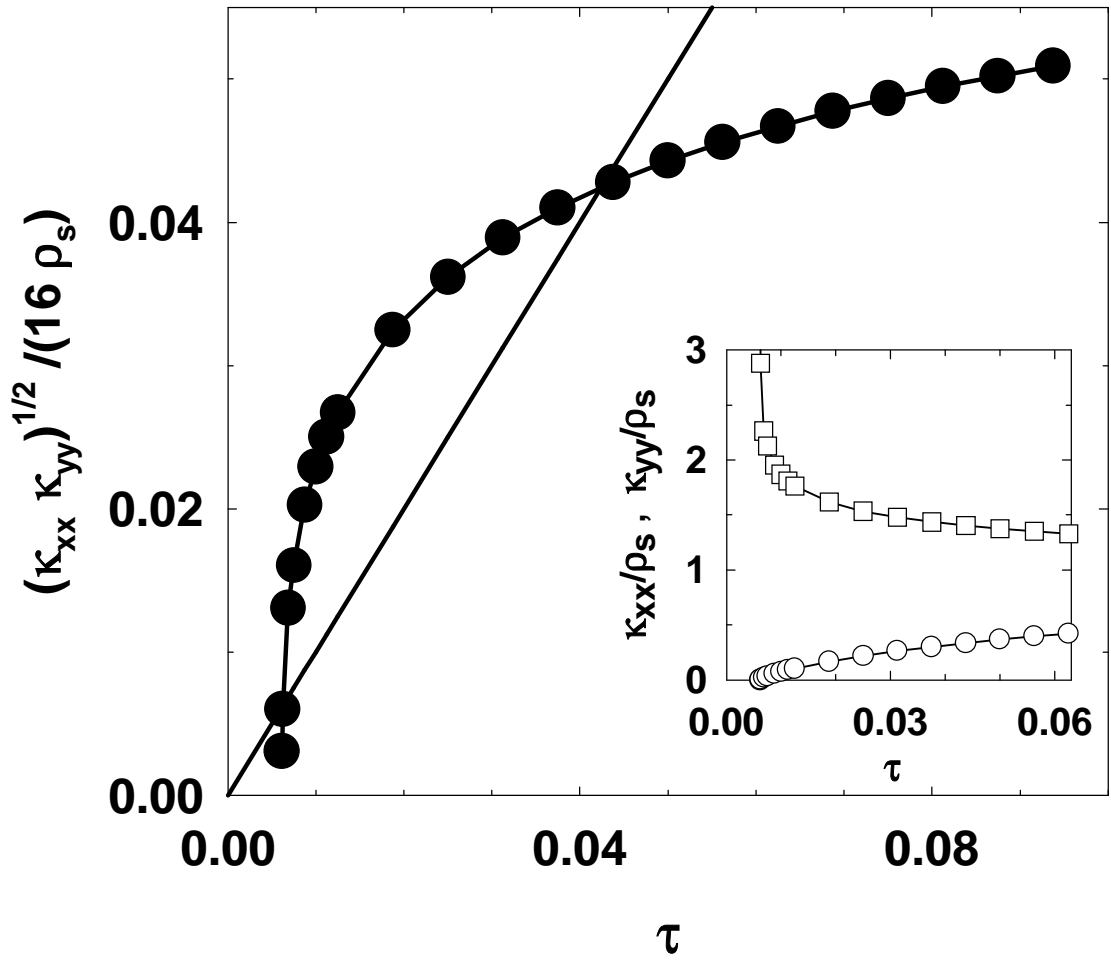
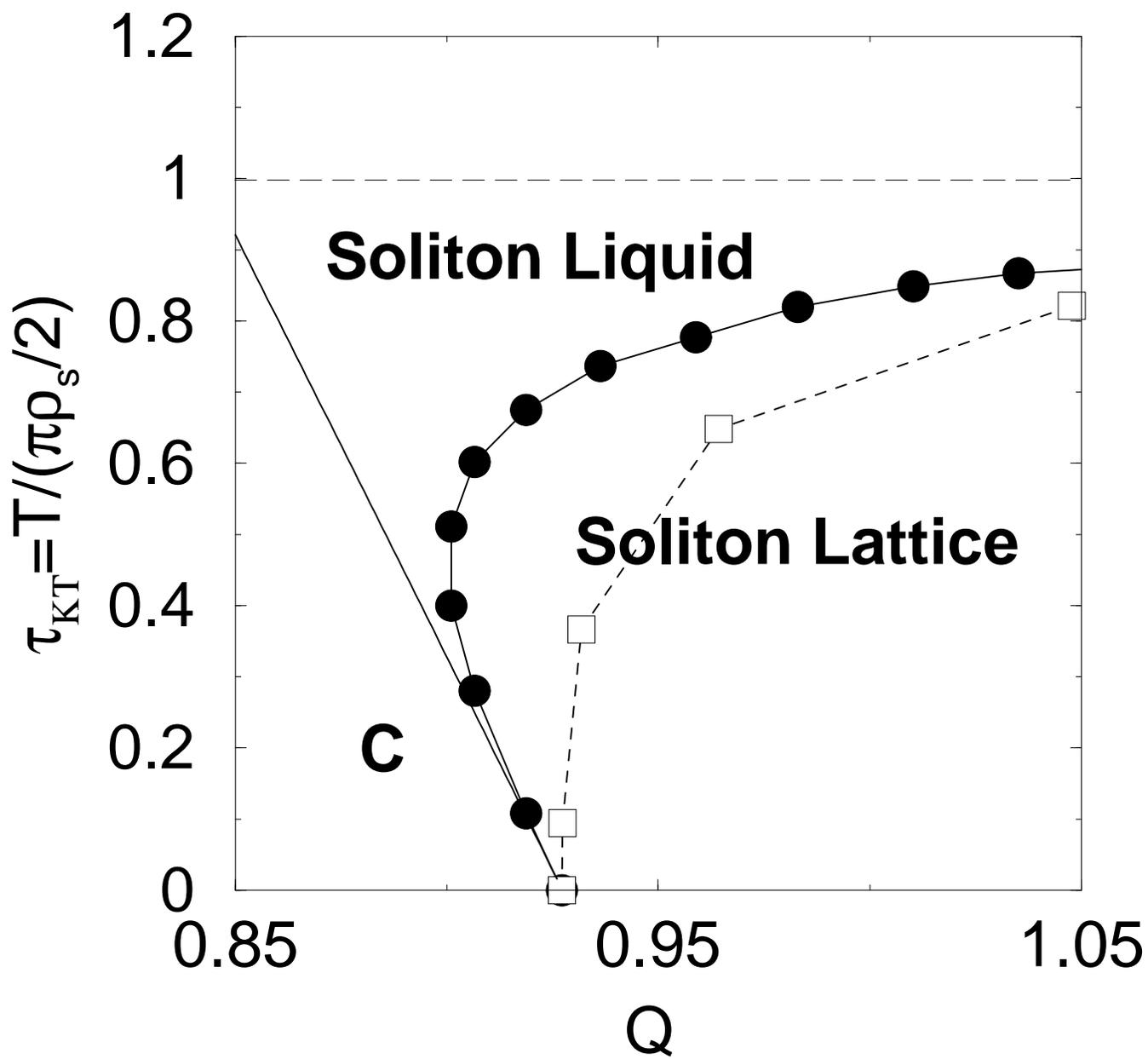
\end{document}